\RequirePackage{ifpdf}
\documentclass[11pt,a4paper]{JHEP3}
\usepackage{amssymb,epsfig,amsmath,slashed,cite}
\usepackage{bm}
\usepackage{lineno}

\newcommand{\no}{\nonumber \\}
\newcommand{\gsim}{\mathrel{\hbox{\rlap{\lower.55ex \hbox {$\sim$}}
                   \kern-.3em \raise.4ex \hbox{$>$}}}}
\newcommand{\lsim}{\mathrel{\hbox{\rlap{\lower.55ex \hbox {$\sim$}}
                   \kern-.3em \raise.4ex \hbox{$<$}}}}

\def\be{\begin{equation}}
\def\ee{\end{equation}}
\def \ba{\begin{equation}}
\def\ea{\end{equation}}
\def \bes {\begin{subequations}}
\def \ees{\end{subequations}}
\def\calO{{\cal O}}

\def \tT{\tilde{T}}
\def\arguu{u;\o, k}
\def\arguo{\o,k}
\def\argutxv{t,\vx}
\def\argutx{t,x}
\def\argutxu{t,x;u}
\def\del{{\partial}}
\def \pd {\partial}
\def\vk{{\bm k}}
\def \vx{\bm x}

\def\vB{{\bm B}}
\def \vxi{\bm \xi}

\def\vj {\bm j}
\def\roughly#1{\mathrel{\raise.3ex\hbox{$#1$\kern-.75em%
\lower1ex\hbox{$\sim$}}}}
\def\lsim{\roughly<}
\def\gsim{\roughly>}

\def\({\left(}
\def\){\right)}
\def\[{\left[}
\def\]{\right]}
\def \<{\langle}
\def \>{\rangle}
\def \sph {\rm sph}
\def \CS {\rm CS}
\def \dyn {\rm dyn}
\def \cell {\rm cell}
\def \holo {\rm holo}

\def\Cone {M}
\def\CME{\rm CME}

\def\af{\alpha}

\def\gm{\gamma}
\def\Gm{\Gamma}
\def\ka{\kappa}
\def\k{\kappa}
\def\lam{\lambda}
\def\l{\lambda}

\def\dlt{\delta}
\def\Dlt{\Delta}
\def\eps{\epsilon}

\def\Sig{\Sigma}
\def\sig{\sigma}
\def\s {\sigma}

\def\omg{\omega}
\def \o {\omega}
\def\Omg{\Omega}
\def\tht{\theta}

\def\wg{\wedge}

\title{
The anomalous transport of axial charge:
topological vs non-topological fluctuations 
}

\author{Ioannis Iatrakis,$^1$\footnotemark[1]\, Shu Lin,$^2$\footnotemark[2]\, and Yi Yin$^3$\footnotemark[3]
\\
$^1$Department of Physics and Astronomy, Stony Brook University, Stony Brook, New York 11794-3800, USA
\\
$^2$RIKEN-BNL Research Center, Brookhaven National Laboratory, Upton, New York 11973-5000, USA
\\
$^3$Physics Department, Brookhaven National Laboratory, Upton, New York 11973-5000, USA}

\footnotetext[1]{E-mail address: \email{ioannis.iatrakis@stonybrook.edu}}
\footnotetext[2]{E-mail addresses: \email{slin@quark.phy.bnl.gov}}
\footnotetext[3]{E-mail address: \email{yyin@quark.phy.bnl.gov}}

\abstract{
Axial charge imbalance is an essential ingredient in novel effects
associated with chiral anomaly such as chiral magnetic effects (CME). 
In a non-Abelian plasma with chiral fermions, 
local axial charge can be generated  a) by topological fluctuations
which would create domains with non-zero winding number b) by
conventional non-topological thermal fluctuations. 
We provide a holographic evaluations of medium's response to
dynamically generated axial charge density in hydrodynamic limit and examine if medium's
response depends on the microscopic origins of axial charge imbalance.  
We show a local domain with non-zero winding number would induce a
non-dissipative axial current due to chiral anomaly.
We illustrate holographically that a local axial charge imbalance
would be damped out with the damping rate related to Chern-Simon
diffusive constant. 
By computing chiral magnetic current in the presence of dynamically
generated axial charge density,
we found that the ratio of CME current over the
axial charge density is independent of the origin of axial charge
imbalance in low frequency and momentum limit.
Finally, a stochastic hydrodynamic equation of the axial charge is
formulated by including both types of fluctuations. 
}

\preprint{RBRC-1142}

\begin{document}


\newpage

\section{Introduction
\label{sec:intro}
}
The parity-odd response of a medium with chiral fermions and its deep
relationship to topology and quantum anomalies have attracted
significant interest.
One such effect under extensive study is 
the chiral magnetic
effect (CME)\cite{Kharzeev:2004ey,Kharzeev:2007tn,Kharzeev:2007jp,Fukushima:2008xe}
, 
which is the appearance of a vector current along the direction of an
external magnetic field in the presence of axial
charge imbalance (see Refs.~\cite{Kharzeev:2013ffa, Kharzeev:2015kna}
for a recent review). 
The CME has been demonstrated in various theoretical frameworks, 
such as in hydrodynamics\cite{Son:2009tf,Neiman:2010zi,Bhattacharya:2011tra,Lin:2011mr,Neiman:2011mj,Jensen:2012jh,Jensen:2012jy},
kinetic theories\cite{Loganayagam:2012pz,Son:2012wh,Stephanov:2012ki,Son:2012zy,Gao:2012ix,Chen:2012ca},
perturbative theories\cite{Fukushima:2008xe,Hou:2011ze,Yee:2014dxa},
effective theories\cite{Lublinsky:2009wr,Lin:2011aa,Nair:2011mk,Capasso:2013jva,Jensen:2013vta}
and in the AdS/CFT correspondence
\cite{Rebhan:2009vc,Gorsky:2010xu,Rubakov:2010qi,Gynther:2010ed,Kalaydzhyan:2011vx,Hoyos:2011us}.
A closely related effect is the chiral vortical effects (CVE), which
is the appearance of a current along the direction of vortivity.
CVE and its relation to mixed anomalies has been studied in
\cite{Landsteiner:2011cp,Landsteiner:2011iq,Jensen:2012kj,Hou:2012xg,Golkar:2012kb}. Interesting
properties of chiral media have been discussed in \cite{Khaidukov:2013sja,Avdoshkin:2014gpa,Kirilin:2013fqa}.

Those anomalous effects are not only theoretically
well-motivated,
but also phenomenologically important. 
In a heavy ion collision, 
a very strong magnetic field,
on the order of $eB\sim
m^{2}_{\pi}$,
 is created from the incoming nuclei that are positively charged and move at nearly the
speed of light.
Therefore, CME will convert axial charge fluctuations generated in
heavy-ion collisions into (vector) charge-dependent correlation which
could be potentially detected by experimental observables.
Recently, there have been significant experimental efforts in
searching for CME and other anomalous transport effects (see \cite{Liao:2014ava}
for a review) in heavy ion collision
experiments \cite{Kharzeev:2010gr,Adamczyk:2014mzf,Abelev:2009ac,Abelev:2012pa}.  

One essential ingredient in those anomaly-related effects is the presence of axial charge imbalance. 
For example,
in terms of chiral charge imbalance parametrized by the axial chemical potential $\mu_{A}$,
CME can be expressed as:
\be
\label{CME}
\vj^{\CME}_V=
C_{A}\mu_A e\vB\, , 
\qquad
C_{A}=
\frac{N_c }{2\pi^{2}}\, . 
\ee
Previously, most studies were based on introducing axial charge asymmetry by
hand,  after which the response of the medium to a magnetic field is investigated (see Ref.~\cite{Fukushima:2010vw,Manuel:2015zpa} for exceptional cases). 
%
%
However,
axial charge density $n_{A}(t,\vx)$ is a local and dynamical quantity depending on space and time. 
The medium's response to time-dependent, in-homogeneous axial
charge density and the connections of this response to anomaly have been rarely studied before. 
One motivation of this paper is to fill this gap.

A distinctive feature of local axial charge density (in contrast to
vector charge density) is 
that there are microscopically two different mechanisms for local
generation of axial charge imbalance. 
The first one is
by \textit{topological fluctuations} of gluonic fields which would
create domains with non-zero winding number.
The resulting topological charge will in turn convert into axial charge density via anomaly relation:
\be 
\label{anomaly}
\pd_{\mu}j^{\mu}_{A}
=-2q\, ,
\qquad
q\equiv\frac{g^2}{32\pi^2}\eps^{\mu\nu\rho\sig}\text{Tr}G_{\mu\nu}G_{\rho\sig}\, ,
\ee
where $q$ is topological charge density (Pontryagin density) and $G_{\mu\nu}$ denotes field
strength of gluonic fields.
Such topological fluctuations would in general create both global axial
charge and local axial charge imbalance. 
The second mechanism is through \textit{thermal fluctuations}.
In the absence of topological fluctuations, 
while the whole system is a grand canonical ensemble
(of axial charge), 
each fluid cell of that system can be
considered as a canonical ensemble (of axial charge). 
Therefore axial charge inside a fluid cell still fluctuates due to thermal
fluctuations.
For a non-Abelian plasma with chiral fermions,
both mechanisms would contribute to local axial charge density fluctuations.
Would response of the medium depend on the way that axial charge
density is generated?

%
%
In this paper,
we will consider a de-confined non-Abelian plasma with chiral fermions at
finite temperature. 
We will begin by studying 
the medium's response to 
the interplay between local axial charge density
$n_{A}(t,\vx)$ and $\vj_{A}(t,\vx)$, $q(t,\vx)$ in long time and long wavelength limit (i.e. in
hydrodynamic regime).
In particular, 
we will examine if the relation among those one point functions depends on the microscopic
 origin of local axial charge density.  
As a basis for this study,
we will work in a top-down holographic model,
namely Sakai-Sugimoto model\cite{Sakai:2004cn,Sakai:2005yt}.
The Sakai-Sugimoto model 
is considered to be close to the large $N_c$ QCD with massless chiral quarks in quenched approximation.
It has been widely applied to study anomaly-related
effects (e.g. \cite{Rebhan:2009vc,Yee:2009vw,Kharzeev:2010gd} ).
To model gluonic fluctuations and implement anomaly relation
\eqref{anomaly},
we will consider the dynamics of $C_7$ Ramond-Ramond field and its
Wess-Zumino coupling to flavor sector.  
As we are working in the long time, long wave length limit,
the results presented in this paper are analytic.

We found axial current $\vj_{A}(t,\vx)$ can be created by
in-homogeneity of topological domains.
It is well known that
in-homogeneity of axial charge densities leads to diffusion:
$\vj_{A}(t,\vx)= -D\nabla n_{A}(t,\vx)$ 
where $D$ is the conventional diffusive constant.
However, if the axial charge density results from a local topological
domain which will be represented by an effective field
``$\theta(\argutxv)$'' in this paper ,
such topological domain will also induce an axial charge density and non-dissipative axial current in addition to diffusive current:
\be
\label{jA_new}
n_{A} = \frac{\Gamma_{\CS}}{T}\theta(\argutx)\, , 
\qquad
\vj^{\rm new}_{A} =
\kappa_{\CS} \nabla \tht\, . 
\ee
Here, 
$\Gamma_{\CS}$ and $\kappa_{\CS}$ is related to the behavior of
retarded Green's function $G^{qq}_{R}(t,\vx) \sim
\<[q(t,\vx), q(0,0)]\>$ in hydrodynamic regime:
\be
\label{GqqR}
G^{qq}_{R}(\o, k) = 
\frac{1}{2}\[- i \frac{\Gamma_{\CS}}{T} \o -\kappa_{\CS} k^2 
\]
\, ,
\ee
and $T$ denotes temperature.
It is worthy noting that $\kappa_{\CS}$ term in \eqref{jA_new} is 
opposite to the direction of diffusive current and non-dissipative. 
One way to understand current \eqref{jA_new} is that a topological
domain also carries kinetic energy which would be transfered to chiral
fermions via anomaly relation \eqref{anomaly}.
In Ref.~\cite{Iatrakis:2014dka} by us,
we have derived \eqref{jA_new} based on a generic setting and
presented a brief verification of \eqref{jA_new} in the Sakai-Sugimoto
model.
We provide more details on this calculation in Sec.~\ref{sec:a2} and
elucidate how axial current \eqref{jA_new} is generated from gravity
side of the duality.

Our calculation in Sec.~\ref{sec:E0} confirms that a local chiral charge imbalance $n_{A}(t,\vx)$
will induced a non-zero $q(t,\vx)$ and they are related by
\be
\label{q_nA}
q(t,\vx ) = \frac{n_{A}(t,\vx)}{2\tau_{\sph}}\, .
\ee
$\tau_{\sph}$ here can be interpreted as the axial charge damping
time.
Indeed, substituting \eqref{q_nA} into \eqref{anomaly},
one has:
$\pd_{\mu}J^{\mu}_{A} = -n_A/\tau_{\sph}$.
Eqs.~\eqref{q_nA} hence implies that a non-zero axial charge
density will eventually be damped out by inducing a non-zero $q$.
Furthermore,
we verify by holographic computations that $\tau_{\sph}$ is related to
Chern-Simon diffusive rate $\Gamma_{\CS}$  and susceptibility $\chi$:
\be
\label{tau_sph}
\tau_{\sph} = \frac{\chi T}{2\Gamma_{\CS}}\, .
\ee
Previously, 
relation \eqref{tau_sph} has been derived based on the standard
fluctuations-dissipation argument\cite{Rubakov:1996vz} (see also Sec.~\ref{sec:hydro}).
Very recently, 
$\tau_{\sph}$ has also been computed numerically in a
bottom-up holographic model\cite{Jimenez-Alba:2015awa}.
To best of our knowledge,
current work is the first direct verification of relation \eqref{tau_sph}
in strong coupling regime.

As we find that the axial current in response to axial charge
density depend on how such axial charge imbalance is generated, 
it is natural to ask if chiral magnetic current \eqref{CME} 
also depends on the origin of axial charge imbalance. 
To be quantitative,
we consider the ratio between CME current and axial charge
density in low frequency, small momentum limit in the presence of
constant magnetic field:
\be
\label{r_CME}
\(\chi_{\dyn}\)^{-1}
\equiv \lim_{\o, k\to 0}
\[
\frac{\mu_{A}(\arguo)}
{n_{A}(\arguo)}
\] \, ,
\qquad
\mu_{A}(\arguo)\equiv \frac{j^{CME}_{V}(\arguo)}{C_{A}eB}\, . 
\ee
For a system with constant axial charge density $n_{A}$,
the ratio $\vj^{CME}_{V}/(CeB)$ equals to axial
chemical potential $\mu_{A}$ due to \eqref{CME}. 
However, if $n_{A}$ is space-time dependent, 
the definition of axial chemical potential $\mu_{A}$ is ambiguous. 
If one takes the ratio $\vj^{CME}_{V}/(CeB)$ as the generalized
definition of axial chemical potential, 
the ratio $n_{A}(\o, k)/\mu_{A}(\o, k)$ can be interpreted as susceptibility.
For this reason, 
we will call $\chi_{\dyn}$ the ``dynamical axial susceptibility''. 

We would like to emphasis that $\chi_{\dyn}$ \eqref{r_CME} is
conceptually different from chiral magnetic
conductivity\cite{Kharzeev:2009pj}
which is the proportionality coefficient of CME current to the
\textit{time-dependent} magnetic field for a medium with 
\textit{homogeneous, time-independent} axial chemical potential.
In \eqref{r_CME} however, 
magnetic field is constant while $n_{A}$ is space-time dependent. 
It is worthy noting that in realistic situations such as quark-gluon
plasma (QGP) created
in heavy-ion collisions, $\chi_{\dyn}$ would be a relevant measure of CME
as in those situations,
the axial charge density is always generated dynamically. 
Previously,
chiral magnetic conductivity has been calculated for plasma in equilibrium at both weak coupling limit
\cite{Kharzeev:2009pj,Hou:2011ze,Yee:2014dxa} and strong coupling
\cite{Yee:2009vw,Gursoy:2014ela,Gursoy:2014boa} and for plasma
out-of-equilibrium \cite{Lin:2013sga}.
However,
we are not aware any existing literature discussing the ``dynamic
axial susceptibility'' $\chi_{\dyn}$ and
its universality.

We have computed $\chi_{\dyn}$ with $n_{A}(\arguo)$ generated by
topological and thermal (non-topological) fluctuations in Sakai-Sugimoto model.
We found such ratio $\chi_{\dyn}$ \eqref{r_CME} is \textit{in-dependent of} the origin of axial
charge imbalance and equals to static susceptibility $\chi$.
Moreover,
we derive a simple analytic expression \eqref{r_CME_holo} relating the
(integration of) gravity metric to  $\chi_{\dyn}$ which applies to
a large class of holographic model.
From such expression,
we obtain a condition on the universality of $\chi_{\dyn}$.

%
%
Having established the fact that axial charge generate by both topological fluctuations
and thermal fluctuations would contribution to CME current,
it is then important to incorporate both fluctuations in the framework of
stochastic hydrodynamics. 
Recently,
there are encouraging progress 
on applying anomalous hydrodynamics to simulate charge seperation
effects\cite{Hirono:2014oda,Yin:2015fca} and chiral magnetic wave effects\cite{Hongo:2013cqa,Yee:2013cya} in heavy-ion collisions. 
In those studies,
axial charge density enters as the initial conditions while the
fluctuations of axial charge density during hydrodynamic
evolution have been neglected.
Motivated by findings in this paper, 
we formulate a stochastic hydrodynamic equation of axial charge density in Sec.~\ref{sec:hydro}.
Such hydrodynamic equation includes stochastic noise from both
topological fluctuations and thermal fluctuations. 
While it is a direct generalization of the general
framework \cite{Kapusta:2011gt,Kovtun:2012rj,Kapusta:2012zb},
to our knowledge,
stochastic equation \eqref{na_hydro_eq} is new in literature.
We hope our theory would be applied to simulate
phenomenology of anomalous transport in the future.

The paper is organized as follows.
We will begin with a brief review
of pertinent ingredients of Sakai-Sugimoto model and realization of
anomaly relation \eqref{anomaly} in Sec.~\ref{sec:SS}.
Sec.~\ref{sec:response} is devoted to studying medium's response to axial
charge density.
The computation of $\chi_{\dyn}$ \eqref{r_CME} is presented in
Sec.~\ref{sec:CME}.
The stochastic hydrodynamic equation for axial charge is formulated in
Sec.~\ref{sec:hydro}.
We conclude in Sec.~\ref{sec:sum}.
%
%
%

\section{
Sakai-Sugimoto model and chiral anomaly
\label{sec:SS}
}

\subsection{
Set-up of the model and realization of anomaly
\label{sec:set_up}
}

%
%

In this paper,
we will work in Sakai-Sugimoto model \cite{Sakai:2004cn,Sakai:2005yt}.
In this model,
the de-confined phase is given by $D4$
black-brane metric, 
which is a warped product of a $5d$ black hole 
 and $S^{1}\times S^{4}$\cite{Witten:1998zw,Aharony:2006da}. 
The $D4$ brane background is given by \cite{Witten:1998zw}:
\be
\label{d4_metric}
ds^2=\(\frac{U}{R}\)^{3/2}\(-f(U)dt^2+d{\vx}^2+dx_4^2\)+\(\frac{R}{U}\)^{3/2}\(U^2d\Omg_4^2+\frac{dU^2}{f(U)}\)
\, ,
\ee
\be
\label{d4_metric2}
F^{\rm RR}_{(4)}=\frac{2\pi N_c\eps_4}{V_4}\, ,
\qquad e^\phi=g_s\(\frac{U}{R}\)^{3/4}\, ,
\qquad R^3=\pi g_sN_cl_s^3\, ,
\qquad f(U)=1-\(\frac{U_T}{U}\)^3\, .
\ee
Here $x_4$ is the coordinates of $S_{1}$ and $\eps_4$ is the
volume form of the four sphere $S_4$.
In addition,
$V_4=8\pi^2/3$ is the volume of $S_{4}$ and
$g_{s}, l_s$ are string coupling and string length respectively. 
The location of the horizon $U_{T}$ is related to the inverse
temperature:
\begin{align}
\label{T_def}
\frac{4\pi}{3}\frac{R^{3/2}}{U_T^{1/2}}=\frac{1}{T}\, .
\end{align}
The periodicity of $x_4$ is given by
\begin{align}
\dlt \tau=2\pi R_4
= \frac{2\pi}{M_{\rm KK}}.
\end{align}
The background \eqref{d4_metric} is stable for $T>1/(2\pi R_4)=M_{KK}/2\pi$
\cite{Aharony:2006da}.
Finally, 
't Hooft coupling $\lam$ is given by:
\be
\label{l_coupling}
\lam =N_{c}\frac{(2\pi)^2 g_sl_s}{2\pi R_4}=2\pi N_{c}g_s l_s M_{KK}\, .
\ee

%
%
To model gluonic fluctuations,
we will consider the dynamics of $C_{7}$ Ramond-Ramond form. 
The kinetic energy of $C_7$ are given by
\begin{align}
\label{RR_kinetic}
S_{RR}=-\frac{(2\pi l_s)^6}{4\pi}\int_{10}dC_7\wg \(*dC_7\). 
\end{align}
In Sakai-Sugimoto model,
right and left handed quarks are introduced by $N_f$ $D8$ branes and
$N_f$ ${\bar D}8$ branes \cite{Sakai:2004cn,Sakai:2005yt}. 
Right-handed (left-handed) $U(1)$ gauge filed $A_{R}$ ($A_{L}$) lives on $D8$
($\bar{D}8$) and is dual to right-handed (left-handed) current
$J^{\mu}_{R}$ ($J^{\mu}_{L}$) on the boundary. 
The $D8$/${\bar D}8$ branes are separated along the $x_4$ direction, 
with $D8$ branes located at $x_4=0$ and ${\bar D}8$ branes located at
$x_4=\pi R_4$. 
In this work will consider $N_f=1$ though generalization to the case
of multi-flavors is straightforward.

The action of bulk gauge field $A_{R,L}$ or its field strength $F_{R,L}$
 is given by the summation of Dirac-Born-Infeld (DBI) term 
\be
\label{d8_action}
S^{R,L}_{DBI}=-\frac{1}{(2\pi)^8l_s^9}\int d^9xe^{-\phi}\sqrt{-det\(g_{MN}+(2\pi\af'\)F^{R,L}_{MN})}
\, , 
\ee
and Wess-Zumino (WZ) term which couples $A_{R,L}$ to Ramond-Ramond form:
\be
\label{WZ_action}
S^{R,L}_{WZ}=\pm\int_{\Sig9}\Sig_q C_{q+1}\wg tre^{\frac{F^{R,L}}{2\pi}}\, .
\ee
We normalize the RR forms $C$ as in \cite{Sakai:2004cn} and use
hermitian worldvolume gauge field $A$. 
The DBI action is identical for $A_{R}$ and $A_{L}$. 
In WZ term,
plus/minus sign is corresponding to $A_{R}$/$A_{L}$ (i.e. $F^{R}_{MN}$/$F^{L}_{MN}$) respectively. 
Axial gauge field and vector gauge field are related to right-handed
and left-handed gauge field by:
\begin{align}
\label{AV_convention}
A= \frac{A^{R}-A^{L}}{2}\, ,
\qquad
V= \frac{A^{R}+A^{L}}{2}\, . 
\end{align}
The total action we will study then becomes:
\be
\label{action}
S = S_{RR} + S_{\rm DBI} + S_{\rm WZ}\, . 
\ee

%
%
It is instructive to show how axial anomaly relation \eqref{anomaly}
is realized in the current holographic model.
Following holographic dictionary, 
the axial current $j^{\mu}_{A}$ is given by the variation of
holographic
on-shell action  $S_{\holo}$ with respect the boundary value of $a_{\mu}\equiv
A_{\mu}(U\to\infty)$ and $q$ is given by the variation of $S_{\holo}$
with respect to $\theta$:
\be
\label{Jq_holo}
j^{\mu}_{A} = \frac{\delta S_{\holo}}{\delta a_{\mu}}\, ,
\qquad
 q=\frac{\dlt S_{\holo}}{\dlt\tht}\, .  
\ee
Here $\theta$ is determined by the holonomy of $C_1$ on the
compactified $x_4$ direction \cite{Witten:1998uka}:
\be
\label{tht_def}
\tht(t,\vx)=\lim_{U\to \infty}\int dx_4(C^{(4)}_1)\, . 
\ee
One may note that the normalization in \eqref{tht_def} is consistent
with the one found by considering action of
multiple probe color branes.

$C_1$ is related to $C_7$ and $A_{R,L}$ field \cite{Green:1996dd}:
\begin{align}\label{c7c1}
dC_1=(2\pi l_s)^6*dC_7
-A^R\wg \(\dlt(x_4-\pi R_4)dx_4\) +A^L\wg \(\dlt(x_4)dx_4\)\, .
\end{align}
It is clear from \eqref{c7c1} that $C_{7}$ is invariant under axial gauge transformation:
\begin{align}
\label{trans}
\dlt_{\Lambda} A^R=d\Lambda\, , 
\qquad
\dlt_{\Lambda} A^L=-d\Lambda\, ,
\qquad
\dlt_{\Lambda} C_1=-\Lambda\dlt(x_4)dx_4-\Lambda\dlt(x_4-\pi R_4)dx_4\, ,
\end{align}
where $\Lambda$ is an arbitrary scalar function.
At boundary and after integrating out $x_{4}$ dependence, 
the axial gauge transformation is reduced to
\begin{equation}
\label{trans_bou}
\dlt_{\Lambda} a_{\mu}(t, \vx)
= \pd_{\mu}\Lambda (t, \vx)\, ,
\qquad
\dlt_{\Lambda}\theta (t, \vx)
= -2\Lambda (t, \vx)\, .
\end{equation}
Since the action \eqref{action} expressed in terms of $C_7$ and field
strength $F_R/F_L$ are manifestly invariant under axial gauge
transformation \eqref{trans},
$S_{\holo}$ would also be invariant under \eqref{trans_bou}.
We therefore have that for an infinitesimal transformation $\delta \Lambda$:
\begin{multline}
\label{S_gauge}
\dlt_{\Lambda} S_{\holo}=\int d^4x
\[\frac{\dlt S_{\holo}}{\dlt
  a_{\mu}(t, \vx)}\pd_{\mu}\(\dlt\Lambda (t, \vx)\)
  +\frac{\dlt S_{\holo}}{\dlt \tht (t, \vx)}(-2\dlt\Lambda (t, \vx))\]\\
=\int d^4x\[ j^{\mu}_{A}(t, \vx) \(\pd_{\mu}\delta\Lambda (t, \vx)\) -2 q (t, \vx)\delta\Lambda (t, \vx)\]
= -\int d^4x\[\pd_{\mu}j^{\mu}_{A}(t, \vx)+2q (t, \vx)\]\delta\Lambda (t, \vx)
=0\, .
\end{multline}
The anomaly relation \eqref{anomaly} then follows from the requirement that
\eqref{S_gauge} holds for arbitrary $\delta\Lambda$,
This is the holographic realization of axial anomaly in the current model.
Realization of axial anomaly in general $Dp/Dq$ brane can be found in \cite{Casero:2007ae}.

\subsection{Fluctuations of bulk fields}
\label{sec:fluctuations}
%
%
We wish to study medium's response to local axial charge
imbalance. 
To model that process in holography,
we need to introduce sources on the boundary. 
Those sources will excite bulk fields which in turn would generate one
point functions such as axial current $j^{\mu}_{A}$ and $q$ on the
boundary. 
As we discussed in the introduction, 
local axial charge imbalance can be generated by a gluonic
configuration with non-zero winding number and by thermal
fluctuations.
Correspondingly,
we will create axial charge imbalance by putting a non-zero
$\theta(\argutxv)$ and by putting a non-zero axial gauge field
$a_{\mu}(\argutxv)$ on the boundary.
In this work,
we will restrict ourselves to the longitudinal fluctuations,
i.e., 
the source on the boundary are $\theta(\argutx),
a_{t}(\argutx),a_{x}(\argutx)$ where $x$ corresponds to the direction
of non-vanishing current 
and sources will only depend on $\argutx$.
Consequently,
non-zero bulk fields are $A_t$, $A_x$, $A_U, C_7$ and they would only
depend on $t, x, U$\footnote{Note that $C_7$ can also depend on $x_4$. 
This can happen when we consider backreaction of $D8$
brane. We will restric ourselves to lowest mode on $S^1$ with no $x_4$
dependence in this work.}.
For later convenience,
we introduce a dimensionless radial coordinate:
\be
u \equiv U/U_{T}\, . 
\ee
We will also rescale all other dimensionful quantities by 
\be
\label{tT_def}
\tT\equiv 
\sqrt{\frac{U_{T}}{R^3}}
= \frac{4\pi}{3}T
= K T\, ,
\qquad
K\equiv\frac{4\pi}{3}\, . 
\ee
where we have used \eqref{T_def}.

We take following ansartz for $C_7$:
\be
\label{C7_B}
 C_7=B_M(\argutxu) \, dx^M\wg d\sig_{12}\wg \epsilon_4
= 
\[ B_{t}(\argutxu)dt + B_{x}(\argutxu)dx+B_{u}(\argutxu)du\]\wg d\sig_{12}\wg \epsilon_4\, . 
\ee
 where $d\sig_{12}=dx_1\wg dx_2$. 
Here, 
$M, N$ run over $t,x,u$.
Adopting the convention
$\eps_{tx_1x_2x_3x_4\tht_1\tht_2\tht_3\tht_4 u}=1$ with
$\tht_i(i=1,2,3,4)$ being angular variables on $S^4$, we obtain the
explicit expressions for $dC_7$ and $*dC_7$:
\bes
\begin{equation}
dC_7=\[
G_{tx}(\argutxu)dt\wg dx+G_{ut}(\argutxu)du\wg dt
+G_{ux}(\argutxu)du\wg dx \]\wg d\sigma_{12}\wg \epsilon_{4}\, ,
\end{equation}
\be
*dC_{7}
=-u^{-1}
\[  
G^{xu}(\argutxu)dt + G^{ut}(\argutxu) dx +G^{tx}(\argutxu) du
\]\wg dx_4\, , 
\ee
\ees
where $G_{MN}=\pd_MB_N-\pd_NB_M$ is the field strength of $B$.
Indices are raised/lowered
by metric \eqref{d4_metric}, i.e. :
\be
g_{tt}(u) = -u^{3/2}f(u)\, , 
\qquad
g_{xx}(u) = u^{3/2}\, , 
\qquad
g_{uu}(u) = u^{-3/2}f(u)^{-1}\, , 
\ee 
and $f(u)=1-u^{-3}$.
Action \eqref{action}, after performing trivial integration
of $x_{4}$,  can then be written in a compact form:
\begin{align}
\label{compact}
S=\int d^4xdu\(-\frac{1}{4}N_c\sqrt{-\gm}F^{MN}F_{MN}-\frac{1}{4}\sqrt{-g}G^{MN}G_{MN}-K\eps^{LMN}B_LF_{MN}\),
\end{align}
where $F_{MN}\equiv\pd_{N}A_{M}-\pd_{M}A_{N}$ 
is the field strength of $A$. 
Three terms in \eqref{compact} are corresponding to $S_{DBI}, S_{RR},
S_{WZ}$ respectively and the sign convention for Levi-Civita symbol is $\eps^{txu}=1$. 
In \eqref{compact}, 
\begin{align}
\sqrt{-g}=
C_{g} u^{-1}
\, .
\qquad
\sqrt{-\gm}=
C_{\gm} u^{5/2}
\, . 
\end{align}
The dimensionless parameter $C_{g},C_{\gm}$ can be expressed in terms of $M_{\rm KK}$
and $T$ as
\be
\label{Cggamma}
C_{g}=\frac{729\pi K^5M^2_{KK}}{4\l^3 T^2}\, ,
\qquad
C_{\gamma} = \frac{2\l T}{27\pi K M_{KK}}\, .
\ee
Varying \eqref{compact} with respect to $A$ and $B$, 
we obtain equation of motion for $G^{MN}, F^{MN}$:
\bes
\label{eom_FG}
\be
\label{eom_G}
\del_M(\sqrt{-g}G^{MN})=K\eps^{QMN}F_{QM}\, , 
\ee
\be
\label{eom_F}
N_c\(\sqrt{-\gm}\del_MF^{MN}\)=K\eps^{QMN}G_{QM}\, . 
\ee
\ees

\subsection{The prescription for computing one point function 
\label{sec:one_point}
}

In this section, 
we will derive the explicit holographic prescription for computing
one point function $n_{A},j_{A}, q$ from definition \eqref{Jq_holo}.
For this purpose,
we need to first obtain holographic on-shell action and then perform
the variation with respect to the sources, $\theta, a_{t}, a_{x}$.
It is therefore more natural to rewrite the holographic action in terms of
$A$ and $C_1$ field instead of $C_7$.  The reason is that $C_1$ is directly related to the boundary source $\theta$ as seen in Eq.(\ref{tht_def}).
From \eqref{c7c1} and \eqref{C7_B},
we have explicitly:
\be
\label{conBM}
 \sqrt{-g} \, \epsilon_{LMN}G^{MN}=2 K \[ \partial_L \Cone +2 A_L\]\, .
\ee
where $\Cone$ is the $x_4$ component of $C_1$, integrated over the
$x_4$ circle:
\be
M
\equiv  \int d x_{4} C^{(4)}_{1}\, . 
\ee
The boundary value of $M$ is $\theta$.
Note that $G^{MN}$ and $\(\del_N\Cone+2A_N\)$ are invariant under both flavor
and $C_7$ gauge transformations \eqref{trans}.
 Using \eqref{conBM},
\eqref{eom_F} can be written  
in terms of $\Cone$ as
\be
\label{eom_Ma}
N_c\del_M\(\sqrt{-\gm}F^{MN}\)=2K^2\sqrt{-g'}\(\del^N\Cone+2A^N\)\,
,
\ee
where we have defined:
\be
\label{gM_def}
\sqrt{-g'} \equiv
\frac{\(-g_{tt}g_{xx}g_{uu}\)}{\sqrt{-g}}\, . 
\ee
Moreover, the Bianchi identity of $G_{MN}$ reads
\be
\label{Bianchi}
\epsilon^{LMN}\pd_{L}G_{MN}=0\, , 
\ee
together with relation \eqref{conBM} gives
\be
\label{eom_Mb}
\del_N\[\sqrt{-g'}\(2A^N+\del^N\Cone\)\]=0.
\ee
It is easy to see the action in terms of the $M$ and $A_M$ fields, which would lead to equation of motion
\eqref{eom_Ma} and \eqref{eom_Mb} is
\begin{align}\label{C1_action}
S=\int
d^4xdu
\bigg[-\frac{1}{4}N_c\sqrt{-\gm}F^{MN}F_{MN}-\frac{K^2}{2}\sqrt{-g'}\(\del_Q\Cone
+2A_Q\)\(\del^Q\Cone +2A^Q\)\bigg] \, ,
\end{align}
as it is analyzed in  \cite{Green:1996dd}, \cite{Casero:2007ae}. The action has the same form as Eq.~(9) in \cite{Jimenez-Alba:2014iia} from a bottom-up model.
 However our action differs in the $N_c$
 dependence of different terms in \eqref{C1_action}, which is absent in
 \cite{Jimenez-Alba:2014iia}. 

In order to compute the one point functions of $q, n_A$ and $j_A$ we use the action (\ref{C1_action}) instead of (\ref{compact}), since it is expressed in terms of the bulk fields which are directly related to the sources of the boundary operators.
To obtain the on-shell action, we do variation of \eqref{C1_action}:
\begin{align}
\dlt S&=\int
d^4xdu\{ \del_M\[ N_c\sqrt{-\gm}F^{MN}-2K^2\sqrt{-g'}\(\del^N\Cone+2A^N\)\]\dlt
A_N
+K^2\sqrt{-g'}\del_N\(\del^NM+2A^N\)\dlt M\} \no
&+\int d^4x\[N_c\sqrt{-\gm}F^{NU}\dlt A_N-K^2\sqrt{-g'}(\del^u\Cone+2A^u)\dlt \Cone\].
\end{align}
The bulk term vanishes by Eqs.~\eqref{eom_Ma}, \eqref{eom_Mb}, 
and the boundary term gives the on-shell action. 
We therefore have from \eqref{Jq_holo}
\be
\label{q_dict}
q(\argutx)=
\lim_{u\to\infty}
\[-K^2\sqrt{-g'}\(\del^u\Cone(\argutxu) + 2A^u(\argutxu)\)\]_{\rm
  Ren}=\lim_{u\to\infty}\[-KG_{tx}(\argutxu)\]_{\rm
  Ren}\, ,
\ee
and
\be \label{njA_dict}
n_A(\argutx)
=N_c\lim_{u\to\infty}\[\sqrt{-\gm}F^{t u}(\argutxu)\]_{\rm Ren}\, ,
\qquad
j_{A}(\argutx)
=N_c\lim_{u\to\infty}\[\sqrt{-\gm}F^{x u}(\argutxu)\]_{\rm Ren}\, .
\ee
In \eqref{njA_dict},
we have introduced a ``bulk axial current''\cite{Iqbal:2008by}:
\be
\label{JA_bulk}
J^{\mu}_{A}(u)\equiv N_{c}\sqrt{-\gamma} F^{\mu u}(u)\, . 
\ee
As in general bulk current $J^{\mu}_{A}(u)$ and bulk field $G_{tx}(u)$
might be divergent near the boundary $u\to \infty$.
we use the subscript ``Ren'' in \eqref{njA_dict} and \eqref{q_dict} to denote the subtraction of such
 divergences in \eqref{q_dict}.
The correspondence: \eqref{njA_dict} and \eqref{q_dict} has been used in
Ref.~\cite{Iatrakis:2014dka}.
It is also interesting to note that from the $u$-component of the first equation in \eqref{eom_G} :
\begin{align}
\label{FU}
N_c\pd_{\mu}\[\sqrt{-\gm}F^{\mu u}\]=2 K G_{tx}\, , 
\end{align}
the anomaly relation \eqref{anomaly} will
be reproduced by taking $u\to\infty$ limit on
both side of \eqref{FU} using \eqref{njA_dict} and \eqref{q_dict}.

We would like to comment on the nature of current $j_A$. 
Naively $j_A$ obtained by a functional derivative is by definition a
consistent current with respect to flavor gauge.
In fact, it is also the covariant current. We can confirm this by
noting that boundary source entering the bulk field strength only
through boundary values of $E_A$ and $G$, which are axial gauge
invariant, therefore $j_A$ is manifestly axial gauge invariant. 
The agreement of consistent and covariant currents may appear
odd: this is because the QCD anomaly studied in this section is
realized with an on-shell action that is manifestly axial
gauge invariant. Therefore, the current obtained from functional
derivative is also invariant, as if it were an ordinary current.
In contrast, the QED anomaly is realized with an anomalous on-shell
action under axial gauge transform. In this case, we can not
have a current which is both conserved (consistent) and invariant
(covariant), in the presence of external axial field. As an example, we
will see that the covariant current is not conserved in
section.~\ref{sec:CME}, where we expand our study to include QED anomaly.
Furthermore, because holography has access to gauge invariant (with respect to $SU(N_{c})$ gauge) quantities only, $j_A$ is also covariant current with respect to $SU(N_c)$ gauge transform.
We stress that the action \eqref{C1_action} is
very different from what we would have obtained by a naive
substitution of \eqref{conBM} into \eqref{compact}. 
In particular, 
the kinetic term of $\Cone$ would have an opposite sign, which would lead to a wrong sign for $q$, \cite{Sakai:2004cn, Green:1996dd, Casero:2007ae}.

%
%

\subsection{Computing one point function
\label{sec:boundary}
}

We now ready to compute one point function.
We will work in Fourier space: $\del_t\to -i\omg$, $\del_x\to ik$. 
Then using \eqref{FU} and Bianchi identity,
one would  express bulk current $J^{t}_{A}, J^{x}_A$ as:
\bes
\label{j_gio}
\be
J^{t}_A(\arguu)=\[\frac{N_c\sqrt{-\gm}(ik)f E'_A(\arguu)-2K (i\omg) G(\arguu)}{\omg^2-k^2f}\]\, ,
\ee
\be
J^{x}_A(\arguu)=\[\frac{N_c\sqrt{-\gm}(i\omg) fE'_A(\arguu)-2K (ik)f G(\arguu)}{\omg^2-k^2f}\]\, .
\ee
\ees
Here we have introduced short-handed notations:
\be
\label{EHG}
E_A(u;\omg,k)\equiv 
-F_{tx}(u;\omg,k)\, ,
\qquad
G(u;\omg,k)\equiv -G_{tx}(u;\omg,k)\, .
\ee
Here and hereafter, we use prime to denote the derivative with respect
to $u$.

From \eqref{j_gio},
we observe that we only need to solve equations for $G(u;\omg, k),
E_{A}(u;\omg,k)$ to obtain one point function $n_{A}, j_{A}, q$.
From \eqref{eom_FG}, 
one finds: 
\bes
\begin{multline}\label{eom_gio}
G''+\(-\frac{1}{u}+\frac{\omg^2f'}{f(\omg^2-k^2f)}\)G'+
\frac{(\omg^2-k^2f)}{u^3f^2}G 
=\\
\(\frac{1}{N_{c}}\)\[\frac{4K^2}{\sqrt{-g}\sqrt{-\gm}}\]G
+\[\frac{2K\omg k f'}{\sqrt{-g}f\(\omg^2-k^2f\)}\]E_{A}\, ,
\end{multline}
\begin{multline}\label{eom_gio2}
E_A''+\(\frac{5}{2u}+\frac{\omg^2f'}{f(\omg^2-k^2f)}\)E_A'+
\frac{(\omg^2-k^2f)}{u^3f^2}E_A  
= \\
\(\frac{1}{N_{c}}\)\[\frac{4K^2}{\sqrt{-g}\sqrt{-\gm}}\]E_{A}
+\(\frac{1}{N_{c}}\)\[\frac{2K\omg k f'}{\sqrt{-\gm}f\(\omg^2-k^2f\)}\]G\, .
\end{multline}
\ees

It is understood that the back-reaction of the flavor branes will
induce $1/N_c$ correction to the black-brane metric. Analysis shows
that the correction to the metric could induce terms $\sim G/N_c$ and
$\sim E_A/N_c$ to \eqref{eom_gio} and terms $\sim E_A/N_c$ and $\sim
G/N_c^2$ to \eqref{eom_gio2}.
we will seek solutions to the leading nontrivial order in power series of $1/N_{c}$:
\be
\label{Nc_expansion}
G=G^{(0)}+\frac{1}{N_{c}}G^{(1)}+\ldots\, ,
\qquad
E_A=E_A^{(0)}+\frac{1}{N_{c}} E_A^{(1)}+\ldots\, .
\ee
To the leading nontrivial order, the solutions 
are not affected by the back-reaction.
Accordingly,
we will compute one point function such as $q, n_{A}, j_{A}$ to first
non-trivial order in $N_{c}$.
From \eqref{eom_FG} and \eqref{Nc_expansion},
we found that $E^{(0)}$ satisfies the homogeneous equation:
\bes
\label{EG_EOM_Nc}
\be
\label{EA0_EOM}
E_A^{(0)}{}''+\(\frac{5}{2u}+\frac{\omg^2f'}{f(\omg^2-k^2f)}\)E_A^{(0)}{}'+\frac{(\omg^2-k^2f)}{u^3f^2}E_A^{(0)}=0
\, , 
\ee
while $G^{(0)}$ satisfies in-homogeneous equation:
\be
\label{G0_EOM}
G^{(0)}{}''+\(-\frac{1}{u}+\frac{\omg^2f'}{f(\omg^2-k^2f)}\)G^{(0)}{}'+\frac{(\omg^2-k^2f)}{u^3f^2}G^{(0)}=
\frac{2K k\omg u f' }{C_{g} f(\omg^2-k^2f)}E_A^{(0)} \, ,
\ee
At order $1/N_{c}$, we further have:
\be
\label{EA1_EOM}
E_A^{(1)}{}''+\(\frac{5}{2u}+\frac{\omg^2f'}{f(\omg^2-k^2f)}\)E_A^{(1)}{}'+\frac{(\omg^2-k^2f)R^3}{u^3f^2}E_A^{(1)}=\
\frac{
  2 K k\omg f'}{C_{\gm}u^{5/2}f(\omg^2-k^2f)}G^{(0)} \, , 
\ee
\ees
and a similar equation for $G^{(1)}$.

Behavior of $G^{(0)}(\arguu)$, $E_A^{(0)}(\arguu)$ and
$E_A^{(1)}(\arguu)$ near boundary can be determined
directly from \eqref{EG_EOM_Nc}:
\bes
\label{GE_asymp}
\be
G^{(0)}(\arguu)=a_2(\arguo)u^2(1+\cdots)+b_0(\arguo)(1+\cdots)\, ,
\ee
\be
E_A^{(0)}(\arguu)=E_0(\arguo)(1+\cdots)+E^{(0)}_1(\arguo)\(u^{-3/2}+\cdots\)\, ,.
\ee
\be
\label{EA1_asymp}
E^{(1)}_{A}(\arguu)=E^{(1)}_1(\arguo)\(u^{-3/2}+\cdots\)\, ,
\ee
\ees
where $\ldots$ denote terms higher order in $1/u$. 
In \eqref{EA1_asymp} we have defined the solution to in-homogeneous
function \eqref{EA1_EOM} $E^{(1)}_{A}$ in
such a way that $E^{(1)}_{A}(u\to\infty)=0$.
$a_{2}$ here is related to $a_{t}, a_{x}, \theta$ by
\begin{align}
\label{sources_rel}
{ a}_2=\frac{K}{2 C_g}\[ 
(\omg^2-k^2)\tht+
2i\omg a_t+2i k a_x\] \, .
\end{align}
In deriving \eqref{sources_rel},
we have used the relation:
\begin{align}
u^{-1}\pd_{u}G
=-\frac{K}{2 C_g}\[ 
f^{-1}\del_t\(\del_t \Cone+2A_t\)-\del_x\(\del_x\Cone+2A_x\)\]\, , 
\end{align}
which can be derived from \eqref{Bianchi} and \eqref{eom_Mb}.
It is useful to note that $E_{0},a_{2}$ are invariant under
transformation \eqref{trans_bou}. 

As usual,
we impose the infalling wave condition at the black hole horizon for
\eqref{EG_EOM_Nc}:
\be
\label{infalling}
\lim_{u\to u_{H}}G^{(0)}(\arguu), E^{(0)}_{A}(\arguu),
E^{(1)}_{A}(\arguu)
\to (u-1)^{-i\o/3}\, .
\ee
Here $u_{H}=1$ denotes the location of horizon.
Consequently,
with given boundary value $a_{2}, E_{0}$,
\eqref{EG_EOM_Nc} can be solved and 
$b_{0}(\arguo), E^{(0)}_{1}(\arguo), E^{(1)}_{1}(\arguo)$  will be
determined from the resulting solutions.
They are related to one point function $q, n_{A},j_{A}$ via
\eqref{j_gio} and definition \eqref{njA_dict},\eqref{q_dict}.
We therefore have
\bes
\label{j_final}
\be
q(\arguo)=K\, b_0(\arguo) \, ,
\ee
\be
n_A(\arguo)=\frac{1}{\omg^2-k^2}
\[-\frac{3C_{\gm}}{2} (i k) \( N_{c} E^{(0)}_1(\arguo)+E^{(1)}_1(\arguo)\)-2K(i\omg)
b_0(\arguo)\]\, ,
\ee
\be
j_A(\arguo)=\frac{1}{\omg^2-k^2}
\[ -\frac{3C_{\gm}}{2}(i\omg )\( N_{c} E^{(0)}_1(\arguo)+E^{(1)}_1(\arguo)\)-2K (ik) b_0(\arguo)\]\, .
\ee
\ees


\section{Medium's response to chiral charge imbalance
\label{sec:response}
}

In section,
we will solve \eqref{EG_EOM_Nc}. with two different boundary
conditions and consider the relation
between $j_{A}, q$ and $n_{A}$.
Physically, 
we would like to use those two different boundary conditions to model
two different mechanisms for the generation of axial charge imbalance. 
In particular, we consider:
 \begin{enumerate}
 \item[Case 1]{\textbf{axial charge imbalance is generated by a domain
    with non-zero winding number.}
To model this situation, 
we set axial gauge field to be zero at boundary, 
i.e., $a_{t},a_{x}=0$ but turn on a non-zero $\theta(\arguo)$.
Consequently, 
boundary condition for \eqref{EG_EOM_Nc} becomes:
\be
\label{case1}
  a_2(\arguo)=\frac{K}{2 C_{g}}\bigg[(\omg^2-k^2)\tht(\arguo) \bigg],
\qquad
E_{0}(\arguo)=0\, . 
\ee
 }
\item[Case 2] {\textbf{axial charge imbalance is generated by non-topological fluctuations.}
To model this situation, 
we instead set $a_2=0$ and consider a non-zero axial electric field on
the boundary:
\be
\label{case2}
  a_2(\arguo)=0,
\qquad
E_{0}(\arguo)\neq 0\, . 
\ee
}
 \end{enumerate}

\subsection{Medium's response to axial charge imbalance generated by
  topological fluctuations
\label{sec:a2}
}

In this section, 
we will study medium's response to axial charge imbalance generated by
topological fluctuations. 
As we discussed previously,
this amounts to solve \eqref{EG_EOM_Nc} with boundary condition
\eqref{case1} and \eqref{EA1_asymp}.
As in this case,
there is no source term for \eqref{EA0_EOM},
 $E^{(0)}(\arguu)=0$ trivially satisfies \eqref{EA0_EOM}
and consequently 
\eqref{G0_EOM} becomes an homogeneous equation:
\begin{align}
\label{G00}
G^{(0)}{}''+\(-\frac{1}{u}+\frac{\omg^2f'}{f(\omg^2-k^2f)}\)G^{(0)}{}'+\frac{(\omg^2-k^2f)}{u^3f^2}G^{(0)}=0\, .
\end{align}

We will seek the in-falling solution for \eqref{G00}
 in hydrodynamic regime $\o,
k\ll 1$.
In this regime, 
the solution can be obtained analytically by first solving \eqref{G00} order by
order in power of $\o, k$ away from horizon and then determining
integration constants by matching with in-falling wave boundary
conditions near the hrozion.
Away from the horizon, we can drop the third term in \eqref{G00} to
obtain the following solution
\be
\label{gh_u}
g_{h}(u) = 
1
 -\(\frac{i\o}{3} \)\[\int^{u}_{u_{H}} du'\(\frac{3 C_{g}(1-s^2 f)}{\sqrt{-g}f} -\frac{1}{u'-1}\)+
\log(u-1)\] \, ,
\ee
where we have defined
\be
s\equiv \frac{k}{\o}\, ,
\ee
to save notations. 
It is easy to check that behavior of \eqref{gh_u} near the horizon $u\to 1$
can be matched to the
infalling wave behavior in small $\o$ limit:
\be
(u-1)^{-i\frac{\o}{3}}
= 1-\frac{i\o}{3}\log(u-1)+{\cal O}(\o^2)\, . 
\ee
It is also worthy mentioning that the integral over $u'$ in
\eqref{gh_u} is convergent as we have explicitly taken the $\log(u-1)$
outside the integral. 

From boundary condition \eqref{case1},
we then fix the normalization of $G^{(0)}$:
\be
\label{G0_sol}
G^{(0)}(\arguu)
= (\frac{ iK\o}{2 C_{g}}\theta) g_{h}(\arguu)\, . 
\ee
Expanding \eqref{G0_sol} near the boundary, we obtain 
\begin{align}
\label{qtheta_sol}
q(\arguo)= 2K b_0(\arguo)=
\(\frac{K^2}{C_{g}}\)
\{
i \o - k^2\lim_{u\to\infty}\[\int^{u}_{u_{H}}du'\frac{C_{g}}{\sqrt{-g}}
-\frac{u^2}{2}\] + {\cal O}(\o^2)
\}\theta(\arguo)\, .
\end{align}
In \eqref{qtheta_sol},
the subtraction is necessary to remove the divergence near the boundary.
As the ratio $-q(\arguo)/\theta(\arguo)$ should be matched to the behavior of retarded Green's function \eqref{GqqR},
we identify $\Gamma_{\CS}, \kappa_{\CS}$ in the present model:
\be
\label{G_CS_SS}
\frac{\Gm_{\CS}}{T}
=\frac{2 K^2}{C_g}=
\frac{2 K^2}{C_g} \tT^3
=\frac{8\l^3 T^6}{729\pi M^2_{KK}}\, , 
\, , 
\ee
\be
\label{K_CS}
\kappa_{\CS} = 
-2\(\frac{K^2}{C_{g}}\)\lim_{u\to\infty}\[\int^{u}_{u_{H}}du'u'
-\frac{u^2}{2}\] 
=
\frac{1}{2}\(\frac{\Gm_{\CS}}{T}\)\tT^{-1}
=\frac{\l^3 T^4}{243\pi^2 M^2_{KK}}\, ,
\ee
where in the last step, we recover the units and used \eqref{Cggamma}
and definition \eqref{tT_def}.
The $\Gamma_{\CS}$ in Sakai-Sugimoto model was computed previously in
\cite{Craps:2012hd}\footnote{our results \eqref{G_CS_SS} has a
  different normalization from \cite{Craps:2012hd}} (see Ref.~\cite{Gursoy:2012bt} for $\Gamma_{\CS}$
in other holographic models). 

To compute $n_{A}, j_{A}$,
one needs to solve in-homogenous equation \eqref{EA1_EOM}:
\be
\label{EA1_EOM2}
E_A^{(1)}{}''+\(\frac{5}{2u}+\frac{\omg^2f'}{f(\omg^2-k^2f)}\)E_A^{(1)}{}'+\frac{(\omg^2-k^2f)R^3}{u^3f^2}E_A^{(1)}=\
\frac{
  2 K k\omg f'}{C_{\gm}u^{5/2}f(\omg^2-k^2f)}G^{(0)} \, , 
\ee
 with $G^{(0)}(\arguu)$ given by \eqref{G0_sol}.
At leading order in $\o, k$, 
the solution reads
\be
\label{EA1_LO}
 E^{(1)}_{A}(\arguu)
=- \(\frac{2 i k  K^2}{3 C_{\gamma}}C_g\)\theta(\arguo) u^{-3/2}\(1+{\cal O}(\arguo)\)\, ,
\ee
which can be easily verified by substituting it into \eqref{EA1_EOM} and
comparing results at leading order in $\arguo$.
Now substituting \eqref{G0_sol} and \eqref{EA1_LO} into \eqref{j_final},
we have obtained the axial charge density generated by $\theta$,
\be
\label{nA_1}
n_{A}=\(\frac{2K^2}{C_{g}}\)\theta(\arguo)\[1 + \calO(\o, k)\]
=\(\frac{\Gamma_{\CS}}{T}\)\theta(\arguo)\, ,
\ee
and $j_{A}$ vanishes at this order.
To obtain $j_{A}$, 
we consider $t,x$ components of \eqref{eom_F} in the presence of $G^{(0)}$
given by \eqref{G0_sol}.
We then obtain flow equations of bulk current $J^{t}_{A}(\arguu),
J^{x}_{A}(\arguu)$ along radial direction $u$:
\be
\pd_{u}J^{t}_{A} =-2K G^{(0)}_{ux}+\(i k \sqrt{-\gm}g^{tt}g^{xx} \)
E^{(1)}_{A}\, ,
\qquad
\pd_{u}J^{x}_{A} =2K G^{(0)}_{ut}+\(i \o \sqrt{-\gm}g^{tt}g^{xx} \)
E^{(1)}_{A}\, . 
\ee
Now using the relation between $G^{(0)}_{tx}$ and $G^{(0)}_{ux},
G^{(0)}_{ut}$ which can be obtained from \eqref{eom_G} in the absence
of $F$,
\be
\label{G_relation}
G^{(0)}_{ut}=\frac{-i kf\pd_{u}G^{(0)}_{tx}}
{\o^2-k^2f}\, ,
\qquad
 G^{(0)}_{ux}=
\frac{i \o\pd_{u}G^{(0)}_{tx}}
{\o^2-k^2f}\, .
\ee
we have from \eqref{G0_sol}:
\bes
\label{njA_flow}
\be
\pd_{u}J^{t}_{A} = - \frac{2
  i\o K^2 \theta}{\sqrt{-g}f}+\(i k \sqrt{-\gm}g^{tt}g^{xx} \) E^{(1)}_{A}
= \frac{- i\o K^2 \theta}{\sqrt{-g}f}\[1+\calO(\arguo)\]\, . 
\ee
\be
\pd_{u}J^{x}_{A} = - \frac{2
  i k K^2 \theta}{\sqrt{-g}}+\(i \o \sqrt{-\gm}g^{tt}g^{xx} \) E^{(1)}_{A}
=- \frac{2
  i k K^2 \theta}{\sqrt{-g}} \[1+\calO(\arguo)\]\, ,
\ee
\ees
By integrating over $u$, we therefore have:
\be
\label{nJA_H}
n_{A}(\arguo) = n^{H}_{A}(\arguo) +\Delta n_A(\arguo)\, , 
\qquad
j_{A}(\arguo) = j^{H}_{A}(\arguo) + \Delta j_{A}(\arguo)\, . 
\ee
Here $n^{H}_{A}, j^{H}_{A}$ are values of bulk current $J^{t}_{A}(\arguu),
J^{x}_{A}(\arguu)$ at horizon $u=u_{H}$. 
We already know $n_A\sim\calO(1)$ from \eqref{nA_1} and 
\be
\Dlt n_A=(-2 i \o K^2\theta)\[\int^{\infty}_{u_{H}}d u' \frac{1}{\sqrt{-g}f}\]\sim\calO(\arguo).
\ee
Therefore we must have
\be\label{nHA}
n_A^H=\(\frac{2K^2}{C_{g}}\)\theta(\arguo)\[1 + \calO(\o, k)\].
\ee
On the other hand,
\be
\label{d_jA}
\Delta j_{A} = 
(-2 i k K^2\theta)\[\int^{\infty}_{u_{H}}d u' \frac{1}{\sqrt{-g}}\]_{Ren}
=(\frac{-2 i k K^2\theta}{C_{g}})\lim_{u\to\infty}\[\int^{\infty}_{u_{H}} d u' u'-u^2/2 \]\, .
\ee
By comparing \eqref{d_jA} with \eqref{K_CS},
we have
\be
\label{d_jA2}
\Delta j_{A} = \ka_{\CS}  (i k ) \theta(\arguo)\, . 
\ee
The current on the horizon
$j^{H}_{A}$ can be determined by substituting \eqref{EA1_LO} into \eqref{j_gio}
and taking $u\to u_{H}$ limit:
\begin{multline}
\label{jHA}
j^{H}_{A}
=
\lim_{u\to u_{H}}\[J^{x}_A(\arguu)\]=
\lim_{u\to u_{H}}\[\frac{\sqrt{-\gm}(i\omg)f \pd_{u}E^{(1)}_A(\arguu)-2K
  (i k) f G^{(0)}(\arguu)}{\omg^2-k^2f}\]\\
= \lim_{u\to u_{H}}\[\sqrt{-\gm} E^{(1)}_{A}(\arguu)\]
=-\frac{2 K^2}{3 C_{g}} (i k \theta)\, . 
\end{multline}
Here we have used a property of any function satisfying in-falling
wave boundary condition, 
say $Z_{\rm in}(u)$ that
\be
\label{infalling2}
\lim_{u\to u_{H}}\(f \pd_{u}Z_{\rm in}\)
= - i \o \lim_{u\to u_{H}}Z_{\rm in}\, . 
\ee

Comparing \eqref{nHA} and \eqref{jHA},
we found that on the horizon,
$j^{H}_{A}$ and $n^{H}_{A}$ are related by Fick's law:
\be
\label{Fick_1}
j^{H}_{A} = -D\nabla n^{H}_{A}\, . 
\ee
To establish \eqref{Fick_1},
we also used the value of diffusive constant $D$ \eqref{D_holo} in
current model. 

To sum up,
in this subsection,
we have studied axial current in response to axial charge imbalance
created by topological fluctuations. 
To represent a domain with non-zero winding number,
we first turn on a non-zero $\theta(\arguo)$ and found that it will
induce a non-zero $q$ \eqref{qtheta_sol} and consequently a non-zero
axial charge density $n_{A}(\arguo)$.
The axial charge density $n_{A}(\arguo)$ and $\theta(\arguo)$ are
related by \eqref{nA_1}. 
Furthermore,
the induced axial current can be divided into two part. 
The first part is due to the diffusion of $n_{A}$ while the second
part is in the opposite direction to the diffusive current and is
proportional to $\k_{\CS}$,
which quantifies the kinetic energy carried by a topological domain.
We verified relation \eqref{jA_new} as first proposed by us in
Ref.~\cite{Iatrakis:2014dka}.
It is interesting to note that holographically, the diffusive
(dissipative) current coincides with the current on the horizon
\eqref{Fick_1} while the non-dissipative current \eqref{jA_new} is given
by the integration from horizon to the boundary (c.f.~\eqref{d_jA}
and \eqref{d_jA2}). 

\subsection{Medium's response to axial charge fluctuations generated by
  non-topological fluctuations
\label{sec:E0}
}

We now consider medium's response to axial charge imbalance generated
by non-topological fluctuations. 
Following our discussion in Sec.~\ref{sec:boundary} 
 we will solve \eqref{EG_EOM_Nc} with boundary condition
\eqref{case1} and \eqref{EA1_asymp}.
We first need to solve the homogeneous solution \eqref{EA0_EOM}.
Similarly to \eqref{G0_sol},
the infalling wave solution to \eqref{EA0_EOM} reads:
\be
e_{h}(\arguu)
= 
1 -\(\frac{i\o}{3}\)
\[\int^{u}_{u_{H}} du'\(\frac{3 C_{\gamma}(1-s^2 f)}{\sqrt{-\gamma}f} -\frac{1}{u'-1}\)+
\log(u-1)\]
\, . 
\ee
Consequently from boundary condition \eqref{case2},
we have:
\be
\label{E0_sol}
E^{(0)}_{A}(\arguu)
= \[\frac{E_{0}(\arguo)}{c_{E}(\arguo)} \]e_{h}(\arguu)\, , 
\ee
where $c_{E}(\arguo)$ is defined by the value of $e_{h}(\arguu)$ at boundary:
\be\label{cE}
c_{E}(\arguo)\equiv e_{h}(u\to \infty;\arguo)
= 1+ i\o \[ s^2
\int^{\infty}_{u_{H}}du'\frac{C_{\gamma}}{\sqrt{-\gamma}}+{\cal
  O}(1) \]
= 1 + \frac{2i\o}{3}\(s^2 +\calO(1)\) \, . 
\ee
Plug \eqref{E0_sol} into \eqref{j_final}, we obtain:
\be
\label{nj_E0}
j_A(\omg, k)=
 N_cC_{\gamma}\frac{E_{0}(\arguo)}{c_{E}(\arguo)}\, ,
\qquad
n_A(\arguo)=  N_cs\,  C_{\gamma} \frac{E_{0}(\arguo) }{c_{E}(\arguo) }\, .
\ee
The conductivity $\s$, diffusive constant $D$ and susceptibility
$\chi$ can be extracted from \eqref{nj_E0} as follows.
First of all,
in the homogeneous limit $s\to 0$ of \eqref{nj_E0},
we will reproduce Ohm's law $j_{A}=\sig E_{A}$.
Therefore:
\be
\label{sigma}
\sig=N_cC_{\gamma}
= \frac{2N_c\l T^{2}}{27\pi M_{KK}}\, .
\ee
On the other hand,
Eqs.~\eqref{nj_E0} must have a hydrodynamic pole corresponding to diffusive
mode at $\o = -i Dk^2$. 
This implies that $c_{E}(\o=-iDk^2,k)=0$ hence:
\be
\label{D_holo}
D = \int^{\infty}_{u_{H}}du'\frac{C_{\gamma}}{\sqrt{-\gamma}} =\
= \int^{\infty}_{u_{H}}du'(u')^{-5/2}
= \frac{2}{3}\tT^{-1}
= \frac{1}{2\pi T}\, .
\ee
In \eqref{sigma} and \eqref{D_holo},
we have used expression \eqref{nj_E0} and have recovered the units at the last step. 
Finally,
using the Einstein relation $\s= \chi D$,
we obtain the expression for $\chi$ from \eqref{sigma} and \eqref{D_holo}:
\be
\label{chi_holo}
\chi^{-1} =\frac{D}{\s}
= \frac{1}{N_c}\int^{\infty}_{u_{H}}\frac{du'}{\sqrt{-\gamma}}\, .
\ee
This relation between $\chi$ and the bulk integration over $\sqrt{-\gamma}$
is in agreement with general expression in Ref.~ \cite{Iqbal:2008by}.


Eq.~\eqref{nj_E0} implies that turning on an external axial electric
field $E_{0}$
would generate a local axial density and axial current.
It would also create a non-zero $q$, which can be determined by solving
\eqref{G0_EOM}:
\be
\label{G0_EOM2}
G^{(0)}{}''+\(-\frac{1}{u}+\frac{\omg^2f'}{f(\omg^2-k^2f)}\)G^{(0)}{}'+\frac{(\omg^2-k^2f)}{u^3f^2}G^{(0)}=
\frac{2K k\omg u f' }{C_{g} f(\omg^2-k^2f)}E_A^{(0)} \, ,
\ee
with $E^{(0)}_{A}$ given by \eqref{E0_sol}.
At leading order in $\o, k$,
the in-homogeneous solution reads:
\be
\label{G0_sol2}
G^{(0)}(\arguu)
=\(\frac{s K}{C_{g}}\) \[
u^2 \(1+ \calO(\arguo)\)+ \frac{2g_{h}(\arguu)}{i\o(1-s^2)}
\]\frac{E_{0}(\arguo)}{c_{E}(\arguo)}\, . 
\ee
As one can check, 
the first term, i.e. , $s K u^2/C_{g}$ term is a special solution to
in-homogeneous equation \eqref{G0_EOM2} at leading order in $\o, k$. 
$g_{h}(u)$ \eqref{gh_u}, the solution to homogeneous equation,
is introduced to guarantee boundary condition \eqref{case2}.
As a result,
we have:
\begin{align}
\label{q_E0}
q(\arguo)=\(\frac{2K^2}{C_{g}}\)\[\frac{-i
  s}{(1-s^2)\omg} \]\frac{E_{0}(\arguo)}{c_{E}(\arguo)}=\(\frac{2K^2}{C_{g}}\)\[\frac{-i
  s}{-s^2\omg} \]\frac{E_{0}(\arguo)}{c_{E}(\arguo)}\, . 
\end{align}
In the last step, we dropped the $1$ in the bracket. This is justified in the diffusive regime where $k^2\sim\o$.
To find the response of $q$ to $n_A$, we first note that $q$ in \eqref{q_E0} contains responses to both $n_A$ and axial electric field $E_0$. In fact, we can exclude the response to the latter by considering $n_A$ induced by a normalizable mode. This occurs when $i\o s^2=-3/2$. According to \eqref{cE} and \eqref{nj_E0}, it implies that $n_A$ remains finite while both $c_E$ and $E_0$ vanish. Physically $n_A$ in this case is induced by a diffusion wave. The response of $q$ to $n_A$ is then given by
\be\label{q_over_nA}
\frac{q}{n_A}=\frac{2K^2}{N_cC_gC_\gm}\frac{1}{-i\o s^2}=\frac{4K^2}{3N_cC_gC_\gm},
\ee
where in the last step we used $i\o s^2=-3/2$ and dropped higher order terms in $\o$. We note that \eqref{q_over_nA} is precisely \eqref{q_nA} as we advocated in the introduction.

We now show that the intuitive argument above could be established more rigourously in real space.
For this purpose,
it is convenient to perform the Fourier transform over $\o$ and
directly consider $n_{A}(t,k), j_{A}(t,k)$ and $q(t, k)$.
By definition and \eqref{nj_E0},
we have:
\begin{multline}
\label{jA_tk}
\frac{j_{A}(t,k)}{N_c}
= \int^{\infty}_{-\infty}\frac{d\o}{2\pi} e^{-i\o t} j_{A}(\arguo)
=-C_{\gamma} \int^{\infty}_{-\infty}\frac{d\o}{2\pi} e^{-i\o
  t}\[\frac{\o E_{0}(\arguo)}{\o+ i D k^2}\] \\
=-C_\gm\int^{\infty}_{-\infty}\frac{d\o}{2\pi} e^{-i\o t}\frac{\o}{\o+iDk^2}\int dt' e^{i\o t'}E_0(t',k).
\end{multline}
The above integral is nonvanishing for $t>t'$ when we can pick up the diffusive pole
in the lower half plane.
We then obtain
\be
\frac{j_{A}(t,k)}{N_c}= -C_\gm Dk^2I_0(t,k).
\ee
$I_{0}(k)$ here is defined by:
\be
I_{0}(k, t)\equiv 
e^{-i D k^2 t} E_{0}(\o=-iDk^2, k)
=
\int^{t}_{-\infty}dt' e^{-Dk^2(t-t')} E(t',k)\, . 
\ee
A similar computation gives:
\be
\label{nA_tk}
\frac{n_{A}(t,k)}{N_c} = (-i C_{\gamma}) k I_0(t, k)\, ,
\ee
\be
\label{q_tk}
q(t, k)= -(\frac{2K^2}{C_{g}})\(-i D k\) I_{0}(t, k)\, . 
\ee
In \eqref{q_tk},
we have neglected a contribution higher order in $k$.

We assume the external field $E_0$ only exists in a finite time window. 
At sufficiently late time $t$, we can regard $j_A$ and $q$ as responses to $n_A$. 
Comparing \eqref{jA_tk} and \eqref{nA_tk} and return to real space,
we obtain Fick's law:
\be
j_{A}(\argutx)= - D\nabla n_{A}(\argutx)\, . 
\ee
Moreover, 
\eqref{nA_tk}, \eqref{q_tk} lead to the relation:
\be
\label{q_nA_holo}
q(\argutx) = \(\frac{2K^2}{N_cC_{g}}\)\(\frac{D}{C_{\gamma}}\) n_{A}(\argutx)
= \(\frac{\Gamma_{\CS}}{T}\)\(\frac{D}{\sigma}\)n_{A}(\argutx)
= \(\frac{\Gamma_{\CS}}{\chi T}\) n_{A}(\argutx)\
= \frac{n_{A}(\argutx)}{2\tau_{\sph}}\, ,
\ee
where we have used \eqref{G_CS_SS}, \eqref{tau_sph}. 

The results of this section can be summarized by the following response matrix, characterizing the response of $q$, $n_A$ and $j_A$ to $\tht$ and $E_A$:
 \begin{align}\label{response_M}
 \begin{pmatrix}
 q\\
 n_A\\
 j_A
 \end{pmatrix}=
 \begin{pmatrix}
 &\frac{i\omg\Gm_\text{CS}}{2T} &\frac{\Gm_\text{CS}}{T}\frac{i\omg}{k(\omg+iDk^2)}\\
 &\frac{\Gm_{\text{CS}}}{T} &\frac{k\sig}{\omg+iDk^2}\\
 &ik\(\ka_{\text{CS}}-D\frac{\Gm_{\text{CS}}}{T}\) &\frac{\omg\sig}{\omg+iDk^2}
 \end{pmatrix}
 \begin{pmatrix}
 \tht\\
 E_0
 \end{pmatrix}.
 \end{align}
We keep only terms lowest order in small $\omg$ and $k$ limit. We further restrict ourselves to the regime $\o\sim k$ for the case with source $\tht$, and to diffusive regime $k^2\sim\o$ for the case with source $E_0$. The transport coefficients we presented are their first nonvanishing order in $N_c$: the responses of $n_A$ and $j_A$ to $E_A$ are at order $O(N_c)$, while the rest responses are at order $O(1)$.

Before closing this section,
we would like to comment on the $O(1)$ correction to the responses of $n_A$ and $j_A$ to $E_0$ above. This requires us to go beyond leading order in $1/N_{c}$ and compute
$E^{(1)}_{A}$ from \eqref{eom_gio2} with $E_A^{(0)}$ given by \eqref{E0_sol}.
Similar analysis shows that near the boundary,
$E_A^{(1)}\sim u^{1/2}+\cdots$ as $u\to\infty$,
which would give divergent contributions to $n_A$ and $j_A$.
This is because the mixing of the bulk fields changes the dimension of the operator.
We note that the change of operator dimension occurs immediately with mixing in bottom-up model in \cite{Jimenez-Alba:2014iia}, while in our case it occurs from the subleading order in $1/N_c$.
As we explained earlier that the solution this order in $1/N_c$ is
incomplete without including back-reaction of the flavor branes. 
It is curious to see if including such backreaction would
remove the potential divergence.
Although these higher order corrections do not affect the results of
our paper, 
we hope that we could revisit the puzzle in future.

\section{Chiral Magnetic Effect and universality
\label{sec:CME}
}

%
%
In the previous section,
we have considered two different situations where axial charge
imbalance is generated.
we now want to study whether the CME current would depend on the microscopic
origin of axial charge density. 
In particular,
we will compute the ratio between CME current $j^{\CME}_{V}$ and
axial charge density $n_{A}$ in low frequency and momentum limit as defined in \eqref{r_CME}.
If axial charge density is static and homogeneous and CME current is
universally given by \eqref{CME},
one would have:
\be
\label{r_chi}
\chi_{\dyn}=\chi\, ,
\ee
due to linearized equation of state $\delta n_{A}= \chi \delta
\mu_{A}$. 
However, 
it is not obvious if \eqref{r_chi} would still hold if $n_{A}$ is generated
dynamically as considered in this paper. 
Of particular interest is the case considered in Sec.~\ref{sec:a2}
that axial charge imbalance is created by topological fluctuations. 


%
%
To compute \eqref{r_CME},
we turn on a small background magnetic field $F^V_{yz}=eB$, 
i.e. magnetic field is longitudinal to the direction of in-homogeneity as
considered in the previous section. 
Then in the presence of bulk axial field $F^A_{tu}$, $F^A_{xu}$ and $F^A_{tx}$,
$F^V_{tu}$, $F^V_{xu}, F^{V}_{tu}$ and $F^V_{tx}$ components of vector field
strength will be excited due to Wess-Zumino term:
\begin{align}
\label{WZ_VA}
S_{\rm WZ}=\int C_3\wg\text{tr}e^{F_R/2\pi}-\int C_3\wg\text{tr}e^{F_L/2\pi}\, ,
\end{align}
with $F_4=dC_3$ given in \eqref{d4_metric}. 
The action for vector field consists of DBI term, which has the same
form as that of axial gauge field $A$ and WZ term from \eqref{WZ_VA}:
\begin{align}\label{action_V}
S_{V}=-N_{c}\int d^4xdU\[
\frac{1}{4}\sqrt{-\gm}F^{MN}_VF^V_{MN}+
K_{B}\eps^{QMN}A_QF^V_{MN}
\] \, ,
\end{align}
where
\be
K_{B}\equiv \(-\frac{C_{A}eB}{2 N_{c}}\)\, . 
\ee
Here, we use subscript/superscript $V$ for vector gauge field strength
$F^V_{MN}$(below we will also use subscript/superscript $A$ for axial gauge field).

The variation of $S_{V}$ gives the equation of motion:
\begin{align}
\label{V_eom}
\del_M\(\sqrt{-\gm}F_V^{MN}\)
=
K_{B}\eps^{NMQ}F^A_{MQ}\, . 
\end{align}
$F^{A}_{MQ}$ will be taken from solutions obtained in previous
solutions. 
To compute \eqref{r_CME},
it is sufficient to work at linear order in $eB$.
We therefore could neglect back-reaction due to $eB$ to the holographic background
and solution $E_{A}$ obtained in the previous section. 

As before, we define 
the bulk vector current
\begin{align}\label{Jv_bulk}
J_V^\mu(\argutxu)=N_c\sqrt{-\gm}F_V^{\mu U}(\argutxu)\, ,
\end{align}
One point functions $n_{V}(\argutx), j_{V}(\argutx)$ are similarly given
by the boundary values of bulk current:
\be
\label{njV_boundary}
n_{V}(\argutx)\equiv\lim_{u\to\infty}J^{t}_{V}(\argutxu)\, ,
\qquad
j_{V}(\argutx)\equiv\lim_{u\to\infty}J^{x}_{V}(\argutxu)\, .
\ee
As before, the vector current defined here is a covariant current.
Similar to \eqref{JA_bulk}, 
it would be convenient to express vector current in terms of $E_V$ and $E_A$
as
\bes
\label{JV_bulk}
\be
J^{t}_V(u)=N_{c}\frac{(ik)\sqrt{-\gm}f E'_{V}(u)-(2i\o
  K_{B})E_{A}(u)}{\omg^2-k^2f(u)}\, ,
\ee
\be
J^{x}_V(u)=N_{c}\frac{(-i\o)\sqrt{-\gm}f\, E'_{V}(u)-(2 i k K_{B})f\,
  E_{A}(u)}{\omg^2-k^2f(u)}\, , 
\ee
\ees
where we have introduced the short-handed notation for ``bulk electric
field'' :
\be
E_{V}(\arguu)\equiv -F^{V}_{tx}(\arguu)\, .
\ee
We can easily verify using \eqref{JV_bulk} that the covariant current
is not conserved in the presence of external axial field $E_A$.
The equation for $E_{V}(\arguu)$ reads
\begin{multline}
\label{Ev_eom}
E_V''+\(\frac{5}{2u}+\frac{\omg^2f'}{\omg^2-k^2f}\)E_V'+\frac{(\omg^2-k^2f)}{u^3f^2}E_V
=\\
\frac{2K_{B}}{C_{\gamma}}\{
\[\frac{\o k f'}{u^{5/2} f\(\o^2-k^2 f\)}\] E_{A}
+\frac{1}{N_{c}}\[\frac{2K \o k f'}{C_{\gamma} u^5 f\(\o^2-k^2 f\)}\] G
\} + {\cal O}\((eB)^2\)\, . 
\end{multline}
We will solve \eqref{Ev_eom} with $E_{A}, G$ determined in the
previous section. 
To concentrate on vector current induced by axial charge imbalance,
we will not turn on any source for vector field, i.e. , imposing
$E_{V}(u\to\infty)=0$ on the boundary and use the standard in-falling
wave boundary condition on the horizon.

To compute the vector current,
it is also convenient to write down ``flow equation'' for bulk vector
current by taking $t, x$ components of \eqref{V_eom} and using definition \eqref{Jv_bulk}:
\bes
\label{njV_flow}
\be
\label{nV_flow}
\pd_{u}J^{t}_{V} (u)= \frac{K_{B}}{\sqrt{-\gamma}f}J^{x}_{A}(u) -\(i k \sqrt{-\gm}g^{tt}g^{xx} \) E_{V} (u)
=\frac{K_{B}}{\sqrt{-\gamma}f}J^{x}_{A}(u)\[1+\calO(\arguo)\]\, . 
\ee
\be
\label{jV_flow}
\pd_{u}J^{x}_{V} (u)=
\frac{2K_{B}}{\sqrt{-\gamma}} J^{t}_{A}(u) -\(i \o \sqrt{-\gm}g^{tt}g^{xx} \) E_{V}(u) 
=\frac{2K_{B}}{\sqrt{-\gamma}} J^{t}_{A}(u)\[1+\calO(\arguo)\]\, ,
\ee
\ees
On the R.H.S of \eqref{jV_flow},
we have used the fact that $J^{t,x}_{V}$ term is always dominated over
$E_{V}$ term in small $\o, k$ limit due to additional gradients in
front of $E_{V}$.
This is because from \eqref{Ev_eom} we observe that $E_{V}$ is the same
order as $E_{A}$ and from \eqref{JA_bulk}, 
$J^{t,x}_{A}$ is at least the same order as $E_{A}$. 

Now integrating \eqref{nV_flow} over $u$ and using definition \eqref{njV_boundary},
we have:
\be
\label{njV_sum}
n_{V}(\arguo) = \Delta n_V(\arguo)+ n^{H}_{V}(\arguo) \, , 
\qquad
j_{V}(\arguo) = \Delta j_{V}(\arguo)+ j^{H}_{V}(\arguo) \, . 
\ee
Here $n^H_V$ and $j^{H}_{V}$ are values of bulk current $J^t_V(\arguu)$ and $
J^{x}_{A}(\arguu)$ at horizon $u=u_{H}$ and 
\be
\Delta n_{V}(\arguo) \equiv 
K_{B}\[\int^{\infty}_{u_{H}}d u' \frac{J^{x}_{A}(\arguu)}{f\sqrt{-\gamma}}\] \, ,
\qquad
\Delta j_{V}(\arguo) \equiv 
2K_{B}\[\int^{\infty}_{u_{H}}d u' \frac{J^{t}_{A}(\arguu)}{\sqrt{-\gamma}}\] \, .
\ee
We now claim that CME current should be identified with $\Delta
j_{V}$, i.e. ,
\be
\label{j_CME}
j^{CME}_{V}
\equiv 2K_{B}\[\int^{\infty}_{u_{H}}d u'
\frac{J^{t}_{A}(\arguu)}{\sqrt{-\gamma}}\] \, .
\ee
The physical motivation behind identification \eqref{j_CME} is that generically in a holographic
set-up, 
the current on the horizon is dissipative (see also example below). 
On the other hand, 
the CME current is non-dissipative.
Therefore one should exclude the horizon current from the total
current when identifying CME current holographically. 

We now consider the implication of \eqref{j_CME}. 
With \eqref{j_CME} and \eqref{chi_holo},
$\chi_{\dyn}$ becomes:
\be
\label{r_CME_holo}
\chi_{\dyn} = 
\lim_{\o, k\to 0}\{
 n_{A}(\arguo)/
\[\int^{\infty}_{u_{H}}d u'
\frac{J^{t}_{A}(\arguu)}{\sqrt{-\gamma}}\]\}\, . 
\ee
It is clear that if in small $\arguo$ limit, bulk axial current is
constant, i.e.,
\be
\label{CME_condition1}
J^{t}_{A}(\arguu) = n_{A}(\arguo) \[1 +{\cal O}(\arguo)\]\, , 
\ee
it follows one can replace $J^{t}_{A}(\arguu)$
with $n_{A}$ in \eqref{r_CME_holo}. 
Consequently,
one will arrive at \eqref{r_chi} by noting \eqref{chi_holo}.
Therefore \eqref{CME_condition1} can be interpreted as a condition for
the validity of \eqref{r_chi}.  

In both cases considered in Sec.~\ref{sec:response}, 
the condition \eqref{CME_condition1} is indeed satisfied, 
we therefor have \eqref{r_chi} for those cases.  

For completeness,
we will calculate total $j_{V}$ for both cases. 
For the first case (c.f.~Sec.~\ref{sec:a2}), 
it is straightforward to check that $j^{H}_{V}\sim {\cal
  O}(\arguo)\theta$,
which is sub-leading compared with $j^{\CME}_{V}$,
we therefore have:
\be
\label{jCME_1}
j_{V}(\arguo)=j^{CME}_{V}(\arguo)
= \(\frac{K_{B}}{\chi}\)\(\frac{T\theta}{\Gamma_{\CS}}\)
= C_{A}\(\frac{\theta}{2\tau_{\sph}}\)eB \, , 
\ee
where we have used \eqref{tau_sph}.
By comparing \eqref{jCME_1} with
\eqref{CME},
it is tempting to make the identification:
\be
\label{muA_theta}
\mu_{A}= \frac{\theta}{2\tau_{\sph}}\, .
\ee
In Ref~\cite{Fukushima:2008xe},
$\mu_{A}$ is identified with $\pd_{t}\theta$.
\eqref{muA_theta} corresponds to replace $\pd_{t}$ in $\pd_{t}\theta$ with the inverse of
characteristic time scale of sphaleron transition $1/\tau_{\sph}$.

The situation is different for the second case (c.f.~ Sec.~\ref{sec:E0}). 
In this case, $j^{H}_{V}$ is the same order as $j^{\CME}_{V}$ in small
$\arguo$ limit. 
From \eqref{nj_E0} and \eqref{r_CME_holo},
we have:
\be
\label{CME_case2}
j^{\CME}_{V}(\arguo)
= s \(\frac{2N_{c}K_{B}}{\chi}\)\(\frac{E_{0}(\arguo)}{c_{E}(\arguo)}\)
= \frac{4 N_{c} K_{B}}{3}\(\frac{k E_{0}(\arguo)}{\o c_{E}(\arguo)}\)\, . 
\ee
On the other hand, to obtain $n_{V}, j_{V}$, 
we need to solve \eqref{Ev_eom}, at leading order in $N_{c}$, that
\begin{multline}
\label{Ev_eom2}
E_V''+\(\frac{5}{2u}+\frac{\omg^2f'}{\omg^2-k^2f}\)E_V'+\frac{(\omg^2-k^2f)}{u^3f^2}E_V
=\frac{2K_{B}\o k f'}{C_{\gamma}u^{5/2} f\(\o^2-k^2 f\)} E_{A}\, ,
\end{multline}
where $E_{A}$ is given by \eqref{E0_sol}.
The leading order solution to \eqref{Ev_eom2} reads:
\begin{align}
\label{Ev_sol2}
E_V=-\frac{4 K_{B} E_{0}(\arguo)}{3c_{E}(\arguo)C_\gm} \[ u^{-3/2}+\calO(\arguo)\]
  s\, .
\end{align}
Again, \eqref{Ev_sol2} can be easily verified by direct
substitution.  
Now substituting \eqref{Ev_sol2} into \eqref{JV_bulk},
we have:
\be
\label{njV_case2}
n_V=-2i N_{c} K_{B}\frac{E_{0}(\arguo)}{\omg c_{E}(\arguo)} \, , 
\ee
and $j_V$ vanishes at this order. Since $\Dlt n_V\sim\calO(1)$, we thus have
\be
n^{H}_{V}(\arguo) =n_{V}(\arguo) = -2i N_{c} K_{B}\frac{E_{0}(\arguo)}{\o c_{E}(\arguo)}\, .
\ee
$j_V^H$ is obtained in the same way as $j_A^H$ in the previous section:
\begin{multline}
\label{jVH}
j^{H}_{V}(\arguo)
=
\lim_{u\to u_{H}}\[J^{x}_V(\arguu)\]=
\lim_{u\to u_{H}}N_c\[\frac{(-i\omg)\sqrt{-\gm}f \pd_{u}E_V(\arguu)-2(i k K_B) 
  f E_A(\arguu)}{\omg^2-k^2f}\]\\
= \lim_{u\to u_{H}}-N_c\[\sqrt{-\gm} E_V(\arguu)\]
=-\frac{4N_{c} K_{B}}{3}\frac{k E_{0}(\arguo)}{\omg
  c_{E}(\arguo)}=-iDk n_V^H(\arguo) \, . 
\end{multline}
Again we see that $j^{H}_{V}$ can be interpreted as a diffusive
current. 
As in this example,
both diffusive current $j^{H}_{V}$ \eqref{jVH} and CME current
\eqref{CME_case2} would contribute to the total vector current in
small $\arguo$ limit, 
to compute CME coefficient hence $\chi_{\dyn}$ properly,
it is crucial to identify CME contribution, i.e., \eqref{j_CME}. 
%
%

To close this section, we would like to comment that while in this paper,
we are working in a specific holographic model,
the relation \eqref{njV_sum} still holds for holographic action for
bulk vector field of the form \eqref{action_V}. 
Consequently, 
assuming the identification of CME current \eqref{j_CME},
the condition \eqref{CME_condition1}  would warrant that
$\chi_{\dyn}=\chi$.
Moreover,
the violation of condition \eqref{CME_condition1} would also
break the relation \eqref{r_chi}.

\section{Stochastic hydrodynamic equations for axial charge density
\label{sec:hydro}
}

We now formulate a hydrodynamic theory for axial charge density by including
stochastic noise from both topological fluctuations and thermal
fluctuations.
We will focus on the dynamics of axial charge thus setting temperature and
fluid velocity $u^{\mu}$ to be homogeneous and time-independent.
We could then work in the frame that the fluid is at rest: $u^{\mu}=(1,0,0,0)$.
To close anomaly:
\be\label{hydro_anom}
\pd_tn_A(t,\vx)+\nabla\cdot\vj_A(t,\vx)=-2q(t,\vx).
\ee
we want to express $q$ and $\vj_{A}$ in terms of noise and $n_{A}$ (or
its gradients).
The constitute relation,
which relates axial current $\vj_{A}$ to $n_{A}$,
is of the conventional form:
\be
\vj_{A}(t,\vx)=
-D \nabla n_{A}(t,\vx) +\vxi(t, \vx) \, , 
\ee
where $\vxi(t,\vx)$ encodes axial charge generated by thermal fluctuations:
\be
\label{vxi}
\<\vxi(t,\vx)\> =0\, , 
\qquad
\<\xi_{i}(t,\vx) \xi_{j}(t,\vx')\>
= 2\sigma T \delta_{ij}\delta(t-t')\delta^3(\vx-\vx')\, . 
\ee
Here $\<\ldots\>$ denotes the average over noise and $i,j=1,2,3$ run over
spatial coordinates.
The magnitude of $\vxi(t,\vx)$ is given by the standard fluctuation-dissipation relation.
Furthermore, $q(t,\vx)$ can be related to $n_{A}(t,\vx)$ using \eqref{q_nA}:
\be
q(t,\vx)= \frac{n_{A}(t,\vx)}{2\tau_{\sph}}+\xi_{q}(t,\vx)\, .
\ee
$\xi_{q}(t,\vx)$ is the noise due to topological fluctuations:
\be
\label{xiq}
\<\xi_{q}(t,\vx)\> =0\, , 
\qquad
\<\xi_{q}(t,\vx) \xi_{q}(t,\vx')\>
= \Gamma_{\CS}\delta(t-t')\delta^3(\vx-\vx')\,, 
\ee
and we will assume that there is no cross correlation between two different
types of fluctuations:
\be
\<\xi_{q}(t,\vx) \xi_{i}(t,\vx')\>
= 0 \, . 
\ee
This completely specifies our stochastic anomalous hydrodynamic equations.
The noise due to topological fluctuations \eqref{xiq} have been considered previously in
Refs.~\cite{Kuzmin:1985mm}.
On the other hand,
for a conserved current (such as vector $\vj_{V}$),
the noise of the form \eqref{vxi} is standard.
Incorporating both fluctuations in \eqref{na_hydro_eq} is new to the
extent of our knowledge.

As an application,
we will consider equal time axial charge correlation
function:
\be
\label{Cnn_def}
C_{nn}(t,\vx)
= \<\[n_{A}(t,\vx)-n_{A}(0,\vx)\]\[n_{A}(t,\vx)-n_{A}(0,\vx)\]\>\, 
. 
\ee
We start with equation for $n_A$ readily followed from
the stochastic hydrodynamic equations 
\eqref{hydro_anom}:
\be
\label{na_hydro_eq}
\[ \pd_{t}  - D\nabla^2 +\tau^{-1}_{\sph}\]n_{A}(t,\vx)
= - \nabla\vxi(t,\vx)+2\xi_{q}(t,\vx) \, .
\ee
Under the initial condition $n_{A}(0,\vx)=0$,
$C_{nn}(t,\vx)$ characterize the magnitude of axial charge fluctuations at
time $t$ and location $\vx$ due to (both) fluctuations.

To compute \eqref{Cnn_def}, 
we first Fourier transform  \eqref{na_hydro_eq} into $\vk$ space (but keep
$t$-dependence):
\be
\label{na_hydro_eqk}
\[ \pd_{t}  + D k^2 +\tau^{-1}_{\sph}\]n_{A}(t,\vk)
= \[-i\vk\cdot\vxi(t,\vk)\]+ 2\xi_{q}(t,\vk) \, .
\ee
The solution of \eqref{na_hydro_eqk} under initial condition
$n_{A}(0,\vx)=0$ reads:
\be
n_{A}(t,\vk)
= 
\int^{t}_{0}dt' e^{-\(Dk^2+\tau^{-1}_{\sph}\)(t-t')}\[(-i
\vk)\cdot\vxi(t,\vk)+2\xi_{q}(t,\vk)\] \, .
\ee
We therefore have:
\begin{multline}
\<n_{A}(t,\vk)n_{A}(t,\vk')\>
= \int^{t}_{0}dt_{1}\int^{t}_{0}dt_{2}\,
e^{-\(Dk^2+\tau^{-1}_{\sph}\)(t-t_{1})-\(Dk'^2+\tau^{-1}_{\sph}\)(t-t_{2})}\\
\times
\[
k_{i}k'_{j}\<\xi_{i}(t_{1},\vk)\xi_{j}(t_{2},\vk')\>+
4\<\xi_{q}(t_{1},\vk)\xi_{q}(t_{2},\vk')\>
\]\, . 
\end{multline}
Using \eqref{xiq} and \eqref{vxi} in Fourier space,
\begin{subequations}
\be
\<\vxi(t,\vk)\> =0\, , 
\qquad
\<\xi_{i}(t,\vk) \xi_{j}(t,\vk')\>
= 2\sigma T \delta_{ij}\delta(t-t')\delta^3(\vk+\vk')\,, 
\ee
\be
\<\xi_{q}(t,\vk)\> =0\, , 
\qquad
\<\xi_{q}(t,\vk) \xi_{q}(t,\vk')\>
= \Gamma_{\CS}\delta(t-t')\delta^3(\vk-\vk')\,, 
\ee
\end{subequations}
and performing the average over noise,
we have:
\begin{multline}
\<n_{A}(t,\vk)n_{A}(t,\vk')\>
= 2\(\sigma T k^2 +2\Gamma_{\CS} \)
\int^{t}_{0} dt_{1} e^{-2 \(Dk^2+\tau^{-1}_{\sph}\)(t-t_1)}\delta^3(\vk+\vk')
\\
=  \frac{\sigma T k^2 +2\Gamma_{\CS} }{Dk^2+\tau^{-1}_{\sph}} 
\[1- e^{-2 \(Dk^2+\tau^{-1}_{\sph}\)t}\]\delta^3(\vk+\vk')
=  \chi T
\[1- e^{-2 \(Dk^2+\tau^{-1}_{\sph}\)t}\]\delta^3(\vk+\vk')\, .
\end{multline}
In the last step, 
we have used Einstein relation $\sigma=\chi D$ and \eqref{tau_sph}.
Now returning to real space,
we have:
\be
\label{Cnn1}
C_{nn}(t,\vx)
= \chi T \int\frac{d^{3}k}{(2\pi)^2}\, 
e^{i\vx\cdot\vk}
 \[1- e^{-2 \(Dk^2+\tau^{-1}_{\sph}\)t}\]
=
\(\chi T\)
\[\delta^3(\vx) -
 \frac{1}{\(8\pi  Dt\)^{3/2}}
e^{-\frac{2t}{\tau_{\sph}}}
e^{-\frac{|\vx |^2}{8 D t}}
\]\, . 
\ee
It is worthy noting that as we are in hydrodynamic regime,
$n_{A}(\vx)$ here should be understood as the coarse-grained axial
charge density inside a fluid cell and $\vx$ is the spatial
coordinates labeling the corresponding fluid cell. 

We now discuss the implication of \eqref{Cnn1}. 
At very early time that $ D t\ll L^{2}_{\cell}$ (therefore $t\ll \tau_{\sph}$) where $L_{\cell}$ is the
size of a fluid cell, 
the Gaussian appearing in \eqref{Cnn1} essentially becomes a delta
function and we then have:
\be
\label{Cnn2}
C_{nn}(t,\vx)
\approx
\chi T
\[1 - e^{-\frac{2t}{\tau_{\sph}}} \]\delta^3(\vx)
\approx
\frac{2 \chi T }{\tau_{\sph}}\, t\, \delta(\vx) 
= 4\Gamma_{\CS}\, t\,\delta^3(\vx)\, ,
\ee
where in the last step we have used \eqref{tau_sph}.
At this stage,
there is no correlation among axial charge in each fluid cell
(c.f.~the delta function in \eqref{Cnn2}).
Integrating \eqref{Cnn2} over volume $\int d^3\vx$,
we further recover relation between the fluctuation of axial charge
and Chern-Simon diffusive constant:
\be
\label{QQ}
\<Q^2_{5}\>
= 4\Gamma_{\CS}V t\, , 
\ee
where $V$ is the volume of the system. 
While it has been widely used in literature to estimate the
fluctuation of axial charge,
\eqref{QQ} is no longer valid at the stage that $D t \sim L^{2}_{\cell}$.
In this stage, 
the spatial dependence of axial charge fluctuations in \eqref{Cnn1} become important.
The diffusion generates additional spatial correlation among
axial charge density.
Finally, 
in the long time limit $t\gg \tau_{\sph} $, 
the second term in \eqref{Cnn1} are suppressed exponentially and axial charge fluctuations
are given by:
\be
\label{Cnn3}
C_{nn}(t\to\infty,\vx)\to 
\(\chi T\)\delta^3(\vx)\, . 
\ee
and 
\be
\label{QQ2}
\lim_{t\to \infty}\<Q^2_{5}\>\to
\chi T V\, , 
\ee
This is of course expected as in long time limit, 
the $\<Q^2_{5}\>$ should approach its thermal equilibrium values
\eqref{QQ2}.

Finally, we remark that to obtain \eqref{QQ2}, 
we have used the relation between Chern-Simon diffusive constant
$\Gamma_{\CS}$ and sphaleron damping rate $\tau_{\sph}$
\eqref{tau_sph}. 
Therefore like Einstein relation $\s=\chi D$ connecting conductivity
$\s$ and diffusive constant $D$, 
the relation between $\Gamma_{\CS}$ and $\tau_{\sph}$ is also fixed by
the requirement based on thermodynamics \eqref{QQ2}. 
It is reassuring that the relation \eqref{tau_sph} is also realized,
as we discussed in Sec.~\ref{sec:E0}, 
in the holographic model studied in this work.

\section{Summary and Outlook
\label{sec:sum}
}

We have analyzed the anomalous transport of a non-Abelian plasma in a
de-confined phase with dynamically generated axial charge using
a top-down holographic model.
In particular, 
we consider two separate cases in which the axial charge is generated
due to a) topological
b) non-topological thermal fluctuations. 
When the axial charge is generated by topological gluonic fluctuations,
we show a non-dissipative current \eqref{jA_new} is induced due to
chiral anomaly in Sec.~\ref{sec:a2}. 
We also illustrate holographically the damping of axial charge due to the
interplay between flavor sector and gluonic sector. 
Furthermore,
we consider the ratio of the CME current to the axial charge density
at small  $\omega$ and $k$ (c.f \eqref{r_CME}).
We interpret such ratio as (the inverse of) ``dynamical axial
susceptibility'' $\chi_{\dyn}$ (c.f.~ the discussion below \eqref{r_CME}).
We found in the context of current holographic model, 
dynamical susceptibility $\chi_{\dyn}$ is independent of the
microscopic origin of the axial charge and coincide with the static
susceptibility $\chi$.
One phenomenological implications of our work, 
in particular,
Sec.~\ref{sec:CME} is that axial charge generated by topological fluctuations and non-topological
thermal fluctuations would contribute to CME signature in heavy-ion
collisions. 
For this reason,
we propose a stochastic hydrodynamic equation of the axial charge
where we incorporate noise both from both fluctuations in Sec.~\ref{sec:hydro}. 
We found that the magnitude of axial charge fluctuations depend on the
time scale that such fluctuations are measured (c.f~\eqref{QQ},
\eqref{QQ2}) and as \eqref{Cnn1} indicates,
the diffusive mode would induce spatial correlations
among axial charges.

There are several issues that can be further studied based on the 
above analysis. The axial charge response to gluonic fluctuations can
be studied when the flavor degrees of freedom  back-react to the glue
part of the theory. In this case, the coupling of the axial current
and the gluonic topological operator, $q$, is not suppressed.  

In this work, 
we found in hydrodynamic limit,
the ``dynamical axial susceptibility'' $\chi_{\dyn}$ \eqref{r_CME} is
universal.
It is both theoretically interesting and phenomenologically important
to extend the definition of $\chi_{\dyn}$ and study its independence
on finite $\omega, k$.
In this case, it is possible that the resulting $\chi_{\dyn}$ would depend on the origin of axial charge imbalance. 

In computing axial charge density correlation function $C_{nn}$ \eqref{Cnn_def}
from stochastic hydrodynamic equation formulated in
Sec.~\ref{sec:hydro}, 
we consider a system in the absence of magnetic field.
Once there is magnetic field, 
axial charge would also be transported by
chiral magnetic wave\cite{Kharzeev:2010gd}. 
Furthermore, 
a new diffusive model would emerge due to the interplay between chiral
magnetic wave and sphaleron damping\cite{Jimenez-Alba:2014iia,Stephanov:2014dma} . 
It is interesting to see how those new modes would contribute to
correlation among axial charge densities within the framework of
stochastic hydrodynamics.

\section*{Acknowledgements}

The authors would like to thank P.~Arnold, U.~Gursoy, C.~Hoyos, A.~Karch, D.~Kharzeev, 
E.~Kiritsis, K.~Landsteiner, H.~Liu, L.~McLerran, G.~Moore, R.~Pisarski, E.~Shuryak, H.-U.~Yee and I.~Zahed for useful
discussions and the Simons Center for Geometry and Physics for
hospitality where part of this work has been done. 
This work is supported in part by the DOE grant No. DE-FG-88ER40388
(I.I.)
and in part by DOE grant No. DE-DE-SC0012704 (Y.Y.) .
S.L. is supported by RIKEN Foreign Postdoctoral Researcher Program.

\bibliographystyle{JHEP}
\bibliography{Q5ref}

\end{document}